\documentclass[a4paper,11pt]{article}
\usepackage{jheppub}
\usepackage[T1]{fontenc} 
\usepackage{enumerate}
\usepackage{graphics}

\title{\boldmath The quantum Fokker-Planck equation of stochastic inflation}
\author[a]{Hael Collins,}
\author[a,b]{Richard Holman,}
\author[a]{and Tereza Vardanyan}
\affiliation[a]{Department of Physics, Carnegie Mellon University\\ 
5000 Forbes Avenue, Pittsburgh, Pennsylvania, U.S.A.}
\affiliation[b]{College of Computational Sciences, Minerva Schools at KGI\\ 
1145 Market Street, San Francisco, California, U.S.A.}
\emailAdd{hcollins@andrew.cmu.edu}
\emailAdd{rh4a@andrew.cmu.edu}
\emailAdd{tvardany@andrew.cmu.edu}

\abstract{We derive the stochastic description of a massless, interacting scalar field in de Sitter space directly from the quantum theory.  This is done by showing that the density matrix for the effective theory of the long wavelength fluctuations of the field obeys a quantum version of the Fokker-Planck equation.  This equation has a simple connection with the standard Fokker-Planck equation of the classical stochastic theory, which can be generalised to any order in perturbation theory.  We illustrate this formalism in detail for the theory of a massless scalar field with a quartic interaction.}


\begin{document}
\maketitle
\flushbottom
\setcounter{page}{2}

\section{Introduction} 

\noindent 
Starobinsky \cite{Starobinsky:1986fx,Starobinsky:1994bd} has long argued that a simple, classical description ought to emerge for a quantum field in an inflating background.  The basis for his description follows from his observation that the coarse-grained theory, containing only the long wavelength fluctuations of a scalar field, satisfies an equation closely resembling the Langevin equation of Brownian motion.  What appears to be happening is that the stochastic forces that act on the long wavelength fluctuations are being generated through their interactions with the short wavelength fluctuations.  Essentially, one part of the field is producing white noise that affects the propagation of the other.  An important result of Starobinsky's approach is that the probability function that determines the classical $n$-point functions of the field is the solution to a simple Fokker-Planck equation.

Despite the compelling simplicity of the stochastic picture, it would appear to be very difficult to see how it could emerge by following the full quantum evolution of the theory.  When we consider a massless, interacting scalar field in a pure de Sitter background, it has been known for even longer that the theory looks pathological \cite{Ford:1984hs,Ford:1985qh,Antoniadis:1985pj,Burgess:2009bs,Seery:2010kh}.  Even in just its free propagation, the free two-point function diverges as it evolves out to late times.  When the loop corrections are treated perturbatively, the degree of their own late-time divergences grows with the number of loops.  Such divergences occur very generically for massless fields in inflating backgrounds, including gauge fields \cite{Prokopec:2007ak,Prokopec:2006ue,Prokopec:2008gw} and the true scalar fluctuations of inflation \cite{Kahya:2010xh}.  Of course, none of these results necessarily contradict Starobinsky's original idea.  Though the basis for the perturbative treatment is the free two-point function, the role of the interactions is obviously crucial.  In fact, the late-time behaviour of even these diverging $n$-point functions is entirely consistent with the predictions of the stochastic picture.\footnote{For example, it is shown in section 4 of \cite{Tsamis:2005hd} that the divergent $n$-point functions from the quantum theory satisfy a recursion relation derived from the stochastic Fokker-Planck equation.}  It has been suspected that through a suitable reorganisation of the perturbative expressions for the $n$-point functions in these theories, a simpler, finite behaviour should emerge that matches directly with the stochastic predictions, for example, in the late-time, static limit.  However, until this has been done, a complete derivation of the stochastic description from the full quantum theory will continue to be lacking.

A recent approach \cite{Burgess:2014eoa,Burgess:2015ajz} to this problem has been to consider the quantum evolution of the theory from a different perspective by working in the Schr\"odinger picture.\footnote{A much earlier treatment of a free scalar field in the Schr\"odinger picture is found in \cite{Mijic:1994vv}.}  In this picture, the connection between the {\it classical\/} probability function of the stochastic description and the {\it quantum\/} density matrix of the scalar theory becomes much clearer.  The essential element that was missing from these works was the fully quantum treatment of the interactions.  In \cite{Burgess:2014eoa,Burgess:2015ajz}, the role of the interactions was only introduced as a background effect on an otherwise quadratic --- purely Gaussian --- theory.

In this article we add this important missing ingredient to show how to derive the stochastic picture for a genuinely interacting field theory.  In particular, we consider here a massless scalar field with a quartic interaction in de Sitter space, and solve for the full time dependence of its density matrix perturbatively in the self-coupling of the field.  Once this evolution has been found to a given order, we can then project onto the theory of the long wavelength fluctuations by integrating out the short wavelength parts of the field.  The resulting density matrix for this effective theory of the long wavelength fluctuations satisfies a fully quantum version of the Fokker-Planck equation.  Essentially, the coarsely grained Liouville equation for the density matrix of the effective theory {\it is\/} the quantum version of the Fokker-Planck equation.  The parallel between the quantum theory and its stochastic description emerges very naturally in this picture.  It becomes a simple matter to read off the stochastic noise and drift from the corresponding quantum version of the equation, as we shall show.

Additionally, by computing the wave-functional for a quartic theory explicitly, we gain a far deeper understanding of the time-evolution of the fluctuations and are able to follow how the structures that depend on the interactions behave both inside and far outside the horizon.  Here we construct the wave-functional for the {\it interacting\/} Bunch-Davies state perturbatively.  When a fluctuation is well inside the horizon, its part of the wave-functional is {\it close to\/} Gaussian; but --- significantly --- it always contains higher-order structures as well.  The role of these higher-order parts grows once the momentum associated with a particular fluctuation crosses the horizon.  We can then see very clearly how their leading behaviour in the long-wavelength limit leads directly to the drift term of the Fokker-Planck equation.

The purpose of this work is not just to verify the validity of the stochastic picture, but to go further and to lay the groundwork for a more powerful formalism.  Having at our disposal a complete derivation that connects the quantum and the classical stochastic theories as we have done here, we can address questions that would be difficult, or otherwise impossible, to approach from the stochastic side.  For example, what are the higher order corrections to the standard stochastic picture?  How does the stochastic limit arise in the theories of other massless fields and what is the influence of their interactions with other fields?  We can even explore, in principle, the degree to which the standard static-limit solution of the stochastic picture is an attractor solution.  Such applications of our approach, together with a few others, are mentioned at the end of this article.

\section{The stochastic description of a quantum theory} 

\noindent 
In a theory of a massless, interacting scalar field, $\Phi(t,\vec x)$, the simplest quantities that we could calculate are the $n$-point functions where all the fields are evaluated at exactly the same space-time point and at some suitably late time,
\begin{equation}
\langle\Phi^n(t,\vec x)\rangle \equiv \lim_{t\to\infty} 
\langle\Omega(t)|\Phi^n(t,\vec x)|\Omega(t)\rangle .
\end{equation}
$|\Omega(t)\rangle$ denotes the state that we have chosen for our quantum field, which we shall take to be the Bunch-Davies state --- the de Sitter invariant state matching the standard Minkowski space vacuum at very short distances.  Because we have chosen a de Sitter invariant state and because we are assuming too that we are working in spatial coordinates where the background is invariant under spatial translations, $\langle\Phi^n(t,\vec x)\rangle$ cannot depend on the position $\vec x$.

Here we are not really interested in the $n$-point functions of the full theory, which contain information about all scales, but only in the $n$-point functions of the effective theory of the long wavelength fluctuations of the field, $\langle\Phi_L^n(t,\vec x)\rangle$.  What we mean by a long or a short wavelength is one whose physical momentum is small or large compared with the Hubble scale, $H$, associated with the curvature of the de Sitter background.  Or, more precisely, we shall use a slightly stricter definition, setting the threshold between `long' and `short' to be well outside the horizon, which can be done by introducing a small parameter $\varepsilon\ll 1$.  
\begin{eqnarray}
\hbox{long wavelength ($L$):} && k<\varepsilon aH 
\nonumber \\
\hbox{short wavelength ($S$):} && k>\varepsilon aH ,
\nonumber 
\end{eqnarray}
$a(t)$ is the scale factor associated with the expanding space-time.  This definition is more appropriate because, with the extremely rapid expansion during inflation, the physical fluctuations corresponding to the scalar fluctuations of inflation which are needed to explain the primordial fluctuations in the early universe would have been stretched far outside the horizon by the end of the inflationary era.  Moreover, if $H$ is meant to be the true cutoff of our effective theory, we should not be including momenta all the way up to this scale.\footnote{In a similar sense one would not use Fermi's theory of $\beta$ decay all the way up to the electroweak scale.}  Of course, in an expanding background the threshold for our effective theory also becomes time-dependent.  If we divide our scalar field $\Phi(t,\vec x)$ into two parts, 
$$
\Phi(t,\vec x) = \Phi_L(t,\vec x) + \Phi_S(t,\vec x) 
= \int_{k<\varepsilon aH} {d^3\vec k\over (2\pi)^3}\, e^{i\vec k\cdot\vec x} \Phi_{\vec k}(t) 
+ \int_{k>\varepsilon aH} {d^3\vec k\over (2\pi)^3}\, e^{i\vec k\cdot\vec x} \Phi_{\vec k}(t) ,
$$
whether a particular $\Phi_{\vec k}(t)$ appears in the first or the second integral depends on the value of $\varepsilon a(t)H$ at that moment.  So `long' and `short' do not have an absolute physical meaning in de Sitter space, but change over time.  A practical consequence --- and one that will later prove to be important in our calculation --- is that this additional time dependence will mean that derivatives can also act on the limits of integrals once we have restricted to just the long-wavelength momenta.

Now suppose that we have determined all of the values of the $n$-point functions of this effective theory.  We could then introduce a classical variable\footnote{Since we are only analysing $n$-point functions whose fields are all evaluated at the same space-time point, a {\it variable\/} $\varphi$ suffices.  If we had wished to consider the expectation values of fields at different points, we should have needed to generalise to a classical stochastic {\it field\/}, $\varphi(\vec x)$, instead.} $\varphi$, together with a probability distribution function $p(t,\varphi)$, such that together they reproduce all the information contained in the functions $\langle\Phi_L^n(t,\vec x)\rangle$.  The weighted average of a power of this variable is defined by the following integral,
\begin{equation}
\langle\varphi^n\rangle(t) 
\equiv \int_{-\infty}^\infty d\varphi\, \varphi^n p(t,\varphi) .
\end{equation}
Notice that while $\varphi$ itself is just a variable without any time dependence, the average $\langle\varphi^n\rangle(t)$ inherits its time dependence from $p(t,\varphi)$.  We can then choose the weighting function $p(t,\varphi)$ so that the expectation values of this classical variable $\varphi$ exactly match with the corresponding $n$-point functions of our effective theory of long wavelength fluctuations,
\begin{equation}
\langle\varphi^n\rangle(t) = \langle\Phi_L^n(t,\vec x)\rangle ,
\end{equation}
once, of course, we have formulated a suitable meaning for $\langle\Phi_L^n(t,\vec x)\rangle$ derived from our original theory.

The stochastic theory of inflation \cite{Starobinsky:1986fx,Starobinsky:1994bd} argues that the probability function for this classical variable should satisfy a Fokker-Planck equation of the form
\begin{equation}
{\partial p\over\partial t} 
= N {\partial^2p\over\partial\varphi^2} 
+ D {\partial\over\partial\varphi}
\biggl( {\partial V\over\partial\varphi} p(t,\varphi) \biggr) .
\end{equation}
The coefficients $N$ and $D$ are called the `noise' and the `drift' of this stochastic theory.  $V(\varphi)$ is a function of the stochastic variable, which is assumed to have the same functional form as the corresponding potential of the quantum theory; that is, one obtains $V(\phi)$ by simply replacing the quantum field $\Phi(t,\vec x)$ with the stochastic variable $\varphi$ in the original quantum potential, 
$$
V(\Phi(t,\vec x)) \xrightarrow{\Phi(t,\vec x)\to\varphi}V(\varphi) .
$$

The fact that $p(t,\varphi)$ is a solution to the Fokker-Planck equation can then be used to generate a recursion relation\footnote{This recursion relation was found already in \cite{Tsamis:2005hd}.} amongst the various averages $\langle\varphi^n\rangle$.  One starts by taking its time derivative,
\begin{equation}
{\partial\over\partial t} \langle\varphi^n\rangle 
= \int_{-\infty}^\infty d\varphi\, \varphi^n\, {\partial p\over\partial t}
= \int_{-\infty}^\infty d\varphi\, \varphi^n\, 
\biggl\{ N {\partial^2p\over\partial\varphi^2} 
+ D {\partial\over\partial\varphi}\biggl( 
{\partial V\over\partial\varphi} p(t, \varphi) \biggr) \biggr\} ,
\end{equation}
and integrates by parts as needed --- twice for the first term and once for the second term --- to produce 
\begin{equation}
{\partial\over\partial t} \langle\varphi^n\rangle 
=  n(n-1) N \langle\varphi^{n-2}\rangle
- nD \biggl\langle \varphi^{n-1} {\partial V\over\partial\varphi} \biggr\rangle .
\end{equation}
For a quartic potential, $V(\varphi)={1\over 4!}\lambda\varphi^4$, this recursion relation has the form
\begin{equation}
{\partial\over\partial t} \langle\varphi^n\rangle
= n(n-1) N \langle\varphi^{n-2}\rangle
- nD{\lambda\over 6} \langle\varphi^{n+2}\rangle .
\end{equation}
So knowing $N$ and $D$, together with the time-dependence of $\langle\varphi^2\rangle$, is sufficient for calculating all the non-vanishing averages, $\langle\varphi^n\rangle$.

What is the origin of this stochastic description of the theory from the perspective of the quantum theory?  The parallels between the stochastic and the quantum versions of the theory emerge most directly when we treat the quantum theory in the Schr\"odinger picture.  In this picture, the evolution of the expectation value of an operator occurs entirely in the state; operators,\footnote{We shall denote the fields in the Schr\"odinger picture with a lower case notation, $\phi(\vec x)$, while the upper case $\Phi(t,\vec x)$ represents the field more generally, independent of a particular picture.} such as the products of $n$ fields, $\phi(\vec x)$, have no explicit time dependence.  The closest analogue of the probability function in the stochastic description is the density matrix --- or rather, its diagonal components --- associated with the state that we have chosen.  In fact we need to treat two versions of the density matrix:  that of the full theory, which we denote by 
\begin{equation}
P[\phi] = P[\phi_L,\phi_S] = \Psi[\phi]\Psi^*[\phi] ,
\end{equation}
as well as the density matrix for the effective theory that just includes the long wavelength fluctuations, which will be denoted by $P_\Omega[\phi_L]$.  The evolution of $P[\phi]$ for the full theory is determined entirely by its Liouville equation.  The evolution of $P_\Omega[\phi_L]$ is then derived through its relation to $P[\phi]$ together with our knowledge of how $P[\phi]$ itself evolves.

A general, equal-time, expectation value for the product of $n$ fields is given by
\begin{eqnarray}
&&\hskip-0.375truein
\langle\Phi_L(t,\vec x_1)\cdots\Phi_L(t,\vec x_n)\rangle 
\nonumber \\
&=& 
\int_L {d^3\vec k_1\over (2\pi)^3} \cdots {d^3\vec k_n\over (2\pi)^3}\,
e^{i\vec k_1\cdot\vec x_1} \cdots e^{i\vec k_n\cdot\vec x_n}\, 
(2\pi)^3 \delta^3(\vec k_1 + \cdots + \vec k_n)\, 
\langle \Phi_{\vec k_1}(t) \cdots \Phi_{\vec k_n}(t) \rangle .
\nonumber 
\end{eqnarray}
The notation that we shall adopt here is that an `$L$' subscript in an integral indicates that all of the momenta accompanying the integral sign are only those corresponding to physical wavelengths that have been stretched well outside the horizon, $k<\varepsilon aH$.  The momentum conserving $\delta$-function follows from the invariance of the background under spatial translations.  When all of the fields are evaluated at the same spatial position, this $\delta$-function causes the exponential factors to vanish,
\begin{equation}
\langle\Phi_L^n(t,\vec x)\rangle 
= \int_L {d^3\vec k_1\over (2\pi)^3} \cdots {d^3\vec k_n\over (2\pi)^3}\, 
(2\pi)^3 \delta^3(\vec k_1 + \cdots + \vec k_n)\, 
\langle \Phi_{\vec k_1}(t) \cdots \Phi_{\vec k_n}(t) \rangle . 
\end{equation}
In the Schr\"odinger picture, the matrix elements are found by functionally integrating over the relevant degrees of freedom, which in this case are the $\phi_{\vec k}$'s whose momentum label $\vec k$ corresponds to a long wavelength, weighted by the density matrix $P_\Omega[\phi_L]$ for the effective theory,
\begin{equation}
\langle \Phi_{\vec k_1}(t) \cdots \Phi_{\vec k_n}(t) \rangle
= \int_L {\cal D}\phi_{\vec k}\,\,\, 
\phi_{\vec k_1} \cdots \phi_{\vec k_n} P_\Omega[\phi_L] . 
\end{equation}
The $\phi_{\vec k}$'s appearing in this expression are the fields written in the Schr\"odinger picture.\footnote{Because of the limits on the integral, the label $\vec k$ is always in the region $k<\varepsilon aH$ in this expression.  It would be redundant --- at least to the order to which we shall be working --- and a little cumbersome to write $\phi_{L,\vec k}$.  Therefore we shall not do so.}  Since the time-dependence is entirely in the density matrix, $\phi_{\vec k}$ does not depend on the time.

Now let us imagine for the moment that the density matrix $P_\Omega[\phi_L]$ itself satisfies a functional Fokker-Planck equation of the form
\begin{equation}
{\partial P_\Omega\over\partial t} 
= \int_L {d^3\vec k\over (2\pi)^3}\, 
\biggl\{ 
{\cal N}_k {\delta^2P_\Omega\over\delta\phi_{\vec k}\delta\phi_{-\vec k}} 
+ {\cal D} {\delta\over\delta\phi_{\vec k}}\biggl[ 
{\delta{\cal V}_\Omega\over\delta\phi_{-\vec k}} P_\Omega \biggr]
\biggr\} . 
\end{equation}
The ${\cal V}_\Omega$ in this expression is the potential for the long wavelength fluctuations of the fields.  For a quartic theory, this potential would be
\begin{equation}
{\cal V}_\Omega[\phi_L] = {1\over 4!}\lambda 
\int_L {d^3\vec k_1\over (2\pi)^3} {d^3\vec k_2\over (2\pi)^3} 
{d^3\vec k_3\over (2\pi)^3} {d^3\vec k_4\over (2\pi)^3}\, 
(2\pi)^3\delta^3(\vec k_1+\vec k_2+\vec k_3+\vec k_4)\,
\phi_{\vec k_1}\phi_{\vec k_2}\phi_{\vec k_3}\phi_{\vec k_4} .
\end{equation}
We can follow the same procedure that we used in the stochastic description of the theory to generate an analogous recursion relation for the quantum effective theory.  When we differentiate $\langle\Phi_L^n(t,\vec x)\rangle$ with respect to the time and use the appropriate quantum form of the Fokker-Planck equation, we are led to the recursion relation\footnote{Had we allowed the drift term in the quantum version of the Fokker-Planck equation to depend on the momentum as well, ${\cal D}_k$, we should have arrived at the following quantum recursion relation instead, 
$$
{\partial\over\partial t} \langle \Phi_L^n(t,\vec x) \rangle
= n(n-1) \biggl( \int_L {d^3\vec k\over (2\pi)^3}\, {\cal N}_k \biggr)
\langle \Phi_L^{n-2}(t,\vec x) \rangle
- n{\lambda\over 6} \langle \Phi_L^{n+2}(t,\vec x) \rangle_{\cal D} , 
$$
where the final $n+2$ point function has been replaced with a `drift-weighted' version of itself,
$$
\langle \Phi_L^{n+2}(t,\vec x) \rangle_{\cal D} \equiv 
\int_L {d^3\vec k_1\over (2\pi)^3} \cdots {d^3\vec k_{n+2}\over (2\pi)^3}\, 
(2\pi)^3 \delta^3(\vec k_1 + \cdots + \vec k_{n+2})\, 
{\cal D}_{|\!| \vec k_n+\vec k_{n+1}+\vec k_{n+2} |\!|} 
\langle \Phi_{\vec k_1}(t) \cdots \Phi_{\vec k_{n+2}}(t) \rangle .
$$
Since the subsequent calculation will show that ${\cal D}$ is momentum independent, we shall simply draw upon this foreknowledge here and not consider this more general possibility for the quantum drift for now.}
\begin{equation}
{\partial\over\partial t} \langle \Phi_L^n(t,\vec x) \rangle
= n(n-1) \biggl( \int_L {d^3\vec k\over (2\pi)^3}\, {\cal N}_k \biggr)
\langle \Phi_L^{n-2}(t,\vec x) \rangle
- n{\cal D}{\lambda\over 6} \langle \Phi_L^{n+2}(t,\vec x) \rangle . 
\end{equation}
This time we have simply written the result for the particular case of a quartic interaction, rather than for a general polynomial potential.  Comparing the two recursion relations, we realise that if the stochastic and the quantum descriptions of the $n$-point functions are to agree, $\langle\varphi^n\rangle = \langle\Phi_L^n(t,\vec x)\rangle$, the noise and the drift coefficients of the stochastic Fokker-Planck equation are derived directly from the quantum ones by identifying 
\begin{equation}
N = \int_L {d^3\vec k\over (2\pi)^3}\, {\cal N}_k 
\qquad\hbox{and}\qquad
D = {\cal D} .
\end{equation}

So the path leading from the quantum theory to its stochastic description now becomes clear:
\vskip-18truept $\quad$

\begin{enumerate}[i.]
\addtolength{\itemsep}{-6pt}
\item We must first solve for the wave-functional and the corresponding density matrix of our full theory, which includes both the long and short wavelength parts of the field.  For this purpose, the Schr\"odinger picture is the best suited, as we shall see.
\item Once we have determined the density matrix for the state that we have chosen, $P[\phi]=P[\phi_L,\phi_S]$, which here will be the Bunch-Davies state, we project onto the effective theory of the long wavelength part of the field.  The most straightforward thing to do is simply to integrate out the short wavelength fluctuations directly and define the density matrix for the effective theory to be
$$
P_\Omega[\phi_L] = \int_S {\cal D}\phi_{\vec p}\, P[\phi_L,\phi_S] .
$$
\item The time dependence of $P_\Omega[\phi_L]$ --- or rather that of the various functions within it --- follows straightforwardly from the time dependence of the functions that appear in the density matrix of the full theory, $P[\phi]$.  Their time dependence, in turn, follows from the Schr\"odinger equation for the wave-functional of the state.
\item Knowing this time dependence of $P_\Omega[\phi_L]$ then allows us to compute its time derivative explicitly.  The resulting equation is a functional Fokker-Planck equation with precisely the form that we claimed that it should have.  This functional Fokker-Planck equation for the effective theory could equally be regarded as the coarse-grained version of the Liouville equation derived from the full density matrix $P[\phi]$.
\end{enumerate}
\vskip3truept

\noindent
We illustrate these steps by applying them to a familiar example.  So we turn next to the case of a scalar field theory with a quartic interaction.

\section{A quartic interaction} 

\noindent 
This method for deriving the stochastic description of a quantum theory is best shown through a particular example.  For this purpose we choose the theory of a real scalar field in a de Sitter background with a quartic self-interaction, $V(\Phi) = {1\over 4!}\lambda\Phi^4$.  Provided that the coupling is sufficiently small, this theory can be solved perturbatively in $\lambda$.  We shall also include a mass for the field for the time being, although we shall ultimately set it to zero.  The action for the theory is written as
\begin{equation}
S[\Phi] = \int dt\, L[\Phi] 
= \int d^4x\, \sqrt{-g}\, \Bigl\{ {\textstyle{1\over 2}}g^{\mu\nu}\partial_\mu\Phi\partial_\nu\Phi
- {\textstyle{1\over 2}} m^2 \Phi^2
- {\textstyle{1\over 24}} \lambda \Phi^4
\Bigr\} .
\end{equation}
The metric $g_{\mu\nu}$ for the de Sitter background can be expressed in a spatially flat form, as we had assumed earlier, either in terms of a `cosmological' time coordinate $t\in(-\infty,\infty)$ or a `conformal' one $\eta\in(-\infty,0)$, 
\begin{equation}
ds^2 = dt^2 - a^2(t)\, \delta_{ij}\, dx^idx^j 
= a^2(\eta)\, \bigl[ d\eta^2 - \delta_{ij}\, dx^idx^j \bigr] .
\end{equation}
We shall later use whichever of these two times best suits our need at the particular moment.  These time coordinates are related to each other through the condition $dt=a(\eta)\, d\eta$, and the scale factor $a$ expressed in these two coordinate systems has the form
\begin{equation}
a(t) = e^{Ht} 
\qquad\hbox{or}\qquad 
a(\eta) = - {1\over H\eta} .  
\end{equation}

Whereas the final result cannot depend on which picture we have chosen, the interaction picture is not the best suited for drawing the parallels between the stochastic and quantum Fokker-Planck equations.  Instead we study the evolution of the theory from a Schr\"odinger perspective.  The time dependence of the state, described in terms of a wave-functional $\Psi[\phi]$, is found by solving the Schr\"odinger equation, 
\begin{equation}
i{\partial\Psi\over\partial t} = H\Psi ,
\end{equation}
where $H[\pi(\vec x),\phi(\vec x)]$ is the Hamiltonian written in terms of the time-independent Schr\"odinger-picture field $\phi(\vec x)$ and its conjugate momentum $\pi(\vec x)$.  In the space-time coordinates that we have chosen, the Lagrangian for our theory is given by 
\begin{equation}
L[\Phi] = \int d^3\vec x\, \Bigl\{ 
{\textstyle{1\over 2}}a^3 \dot\Phi^2 - {\textstyle{1\over 2}}a\delta^{ij} \partial_i\Phi\partial_j\Phi
- {\textstyle{1\over 2}}a^3 m^2 \Phi^2 - {\textstyle{1\over 24}} a^3 \lambda \Phi^4 \Bigr\} ,
\end{equation}
and the corresponding canonical momenta are 
$$
\Pi(t,\vec x) = {\delta L\over\delta\dot\Phi(t,\vec x)} = a^3 \dot\Phi(t,\vec x) . 
$$
If we perform the usual Legendre transformation, we are led to the Hamiltonian,
\begin{equation}
H = \int d^3\vec x\, \bigl\{ \Pi\dot\Phi \bigr\} - L  
= \int d^3\vec x\, \Bigl\{ 
{\textstyle{1\over 2}} a^{-3} \Pi^2 
+ {\textstyle{1\over 2}} a \delta^{ij} \partial_i\Phi\partial_j\Phi 
+ {\textstyle{1\over 2}} a^3 m^2 \Phi^2 + {\textstyle{1\over 24}} a^3 \lambda\Phi^4 
\Bigr\} .
\end{equation}
It is must easier to describe the truncation of the full theory to its long wavelength parts in terms of the momenta of the fields rather than in terms of  their positions.  So, after performing the Fourier transformation of the Schr\"odinger picture fields,
$$
\phi_{\vec k} = \int d^3\vec x\, e^{-i\vec k\cdot\vec x} \phi(\vec x) ,
$$
the Hamiltonian assumes the form
\begin{eqnarray}
H &=& 
\int {d^3k_1\over (2\pi)^3} {d^3k_2\over (2\pi)^3}\, (2\pi)^3\delta^3(\vec k_1+\vec k_2) \biggl\{ 
- {1\over 2} {1\over a^3} {\delta\over\delta\phi_{\vec k_1}}{\delta\over\delta\phi_{\vec k_2}} 
+ {1\over 2} a^3 \biggl( m^2 + {k_1^2\over a^2} \biggr) \phi_{\vec k_1} \phi_{\vec k_2}
\biggr\} 
\nonumber \\
&& 
+ \int {d^3k_1\over (2\pi)^3} {d^3k_2\over (2\pi)^3}{d^3k_3\over (2\pi)^3}{d^3k_4\over (2\pi)^3}\, 
(2\pi)^3\delta^3(\vec k_1+\vec k_2+\vec k_3+\vec k_4) \biggl\{ 
{1\over 24} a^3 \lambda \phi_{\vec k_1} \phi_{\vec k_2} \phi_{\vec k_3} \phi_{\vec k_4} 
\biggr\} .  \qquad\quad
\end{eqnarray}

At this stage, it is not possible to find the exact form of the wave-functional in this interacting theory, so we must be content with constructing $\Psi[\phi]$ perturbatively in powers of the coupling $\lambda$.  One starts by expressing the wave-functional in the form
\begin{equation}
\Psi[\phi] = N e^{-a^3\, \Gamma[\phi]} ,
\end{equation}
where $\Gamma[\phi]$ is a series expanded in powers of the scalar field, $\phi_{\vec k}$,
\begin{equation}
\Gamma[\phi] = \sum_{n=2}^\infty {1\over n!} 
\int {d^3k_1\over (2\pi)^3} \cdots {d^3k_n\over (2\pi)^3}\, 
(2\pi)^3\delta^3(\vec k_1+\cdots +\vec k_n)\,
\Gamma_n(t;\vec k_1,\ldots,\vec k_n)\, \phi_{\vec k_1} \cdots \phi_{\vec k_n} ,
\end{equation}
and $N$ is the normalisation, fixed by the condition,
\begin{equation}
\int {\cal D}\phi_{\vec k}\, \Psi[\phi]\Psi^*[\phi] = 1 .
\end{equation}
The task of solving the Schr\"odinger equation now becomes the problem of determining the detailed form of the functions $\Gamma_n(t;\vec k_1,\ldots,\vec k_n)$.  The fact that the background is invariant under spatial translations in these coordinates has again allowed us to extract a momentum-conserving $\delta$-function.  Furthermore, by a simple relabeling of the momenta over which we are integrating, we can show that these functions are completely symmetric under any permutation of their arguments,
$$
\Gamma_n(t;\vec k_1,\ldots,\vec k_i,\ldots,\vec k_j,\ldots,\vec k_n)
= \Gamma_n(t;\vec k_1,\ldots,\vec k_j,\ldots,\vec k_i,\ldots,\vec k_n) .
$$
When $m^2\ge 0$, the vacuum state of the theory should have the same $\phi\leftrightarrow -\phi$ symmetry as the potential.  This symmetry means that all the odd-order functions vanish,
$$
\Gamma_n(t;\vec k_1,\ldots,\vec k_n)=0
\qquad\hbox{for $n\in\hbox{odd}$.}
$$

To compute the nonvanishing functions, $\Gamma_n(t;\vec k_1,\ldots,\vec k_n)$ with $n\in\hbox{even}$, we need to find the appropriate equations of motion.  This is done by expanding each side of the Schr\"odinger equation in powers of $\phi_{\vec k}$ and matching the terms that share the same numbers of fields.  This process produces a set of coupled differential equations for the functions $\Gamma_n(t;\vec k_1,\ldots,\vec k_n)$.  For example, the left side of the Schr\"odinger equation is evaluated straightforwardly enough,
$$
i {\partial\Psi\over\partial t} 
= i \biggl\{ {\dot N\over N} - a^3 \sum_{n=2}^\infty {1\over n!} 
\int {d^3k_1\over (2\pi)^3} \cdots {d^3k_n\over (2\pi)^3}\, 
(2\pi)^3\delta^3(\vec k_1+\cdots +\vec k_n)\,
\biggl[ {\partial\Gamma_n\over\partial t} + 3{\dot a\over a} \Gamma_n \biggr]\, 
\phi_{\vec k_1} \cdots \phi_{\vec k_n} \biggr\} \Psi ,
$$
but the right side contains a more complicated tower of terms.  These are generated when the functional derivatives in the Hamiltonian act on the wave-functional,
\begin{eqnarray}
H\Psi &=& 
\biggl\{ 
\int {d^3k_1\over (2\pi)^3} {d^3k_2\over (2\pi)^3}\, (2\pi)^3\delta^3(\vec k_1+\vec k_2) \biggl\{ 
{1\over 2} a^3 \biggl( m^2 + {k_1^2\over a^2} \biggr) \phi_{\vec k_1}\phi_{\vec k_2}
\biggr\} 
\nonumber \\
&& 
+ \int {d^3k_1\over (2\pi)^3} {d^3k_2\over (2\pi)^3}{d^3k_3\over (2\pi)^3}{d^3k_4\over (2\pi)^3}\, 
(2\pi)^3\delta^3(\vec k_1+\vec k_2+\vec k_3+\vec k_4) \biggl\{ {1\over 24} a^3 
\lambda \phi_{\vec k_1} \phi_{\vec k_2} \phi_{\vec k_3} \phi_{\vec k_4} 
\biggr\}  
\nonumber \\
&& 
+ \sum_{n=0}^\infty {1\over n!} 
\int {d^3k_1\over (2\pi)^3} \cdots {d^3k_n\over (2\pi)^3}\, 
(2\pi)^3\delta^3(\vec k_1+\cdots +\vec k_n)\,
\nonumber \\
&&\qquad\qquad\times
\biggl[ {1\over 2} \int {d^3k\over (2\pi)^3}\, \Gamma_{n+2}(t;\vec k_1,\ldots,\vec k_n,\vec k,-\vec k) \biggr]\, 
\phi_{\vec k_1} \cdots \phi_{\vec k_n}
\nonumber \\
&& 
+ \sum_{n=0}^\infty \sum_{n'=0}^\infty {1\over n!}{1\over n'!} 
\int {d^3k_1\over (2\pi)^3} \cdots {d^3k_n\over (2\pi)^3}
{d^3k'_1\over (2\pi)^3} \cdots {d^3k'_{n'}\over (2\pi)^3}\, 
(2\pi)^3\delta^3(\vec k_1+\cdots+\vec k_n+\vec k'_1+\cdots+\vec k'_{n'})\,
\nonumber \\
&&\quad
\biggl[ - {a^3\over 2} \Gamma_{n+1}(t;\vec k_1,\ldots,\vec k_n,-\sum \vec k_i) 
\Gamma_{n'+1}(t;\vec k'_1,\ldots,\vec k'_{n'},-\sum \vec k'_i) \biggr]\, 
\phi_{\vec k_1} \cdots \phi_{\vec k_n}\phi_{\vec k'_1} \cdots \phi_{\vec k'_{n'}}
\biggr\} \Psi . 
\nonumber 
\end{eqnarray}
But by collecting and matching the various functions according to the shared factors of $\phi_{\vec k_1} \cdots \phi_{\vec k_n}$ that accompany them for a given $n$, we find a differential equation for each of the $\Gamma_n$'s.  The function $\Gamma_2(t;\vec k,-\vec k)$ accompanying the quadratic part of $\Gamma[\phi]$ obviously depends only on a single momentum.  Since this function occurs ubiquitously throughout the following calculations, it is advantageous to change our notation slightly and write it a little more succinctly as
\begin{equation}
\alpha_k(t) \equiv \Gamma_2(t;\vec k,-\vec k) .
\end{equation}
The Schr\"odinger equation then implies the following relations derived from the zeroth, quadratic, and quartic order terms in the fields,
\begin{eqnarray}
{\dot N\over N} 
&=& 
- {i\over 2} (2\pi)^3\delta^3(\vec 0) \int {d^3\vec p\over (2\pi)^3}\, \alpha_p(t)
\nonumber \\
{\partial\alpha_k\over\partial t} + 3{\dot a\over a} \alpha_k 
&=& 
i\biggl\{ m^2 + {k^2\over a^2} - \alpha_k^2
+ {1\over 2}{1\over a^3} \int {d^3\vec p\over (2\pi)^3}\, \Gamma_4(t;\vec k,-\vec k,\vec p,-\vec p)
\biggr\} , 
\nonumber \\
{\partial\Gamma_4\over\partial t} + 3{\dot a\over a} \Gamma_4(t;\vec k_1,\vec k_2,\vec k_3,\vec k_4) 
&=& 
i\lambda - i \bigl[ \alpha_{k_1} + \alpha_{k_2} + \alpha_{k_3} + \alpha_{k_4} \bigr] 
\Gamma_4(t,\vec k_1,\vec k_2,\vec k_3,\vec k_4)
\nonumber \\
&&
+ {i\over 2}{1\over a^3} \int {d^3\vec p\over (2\pi)^3}\, 
\Gamma_6(t;\vec k_1,\vec k_2,\vec k_3,\vec k_4,\vec p,-\vec p) , 
%\nonumber 
\end{eqnarray}
and so on for yet higher orders of $n$.  The infinite factor $(2\pi)^3\delta^3(\vec 0)$ that appears in the equation for the time dependence of the normalisation, and in several of the equations that will occur later, is the volume of a spatial hypersurface in de Sitter space.  These volume factors always accompany contributions to the normalisation.

It is important to remember that the form of these equations is determined entirely by the dynamical theory that we are considering, that is, by the Hamiltonian of a quartic theory.  Because each of these equations is first-order, there is an additional freedom associated with the choice of the constants of integration\footnote{These are constants in {\it time\/}.  In general they could depend on momenta for particular choices of the state.} appearing in the particular solution for the $\Gamma_n$'s.  The collective choice for all of these constants translates into the choice of a particular state $\Psi[\phi]$ in this picture.

\section{Perturbation theory and the vacuum state} 

\noindent 
The usual stochastic treatment of inflation always implicitly assumes that the theory is in the Bunch-Davies state.  It is therefore important to introduce appropriate conditions on the functions $\Gamma_n$ at very short wavelengths in order to put the field in the correct state.  After we have done so, we can follow the evolution to large wavelengths and see the simplifications that permit a stochastic description of the theory.  We construct the Bunch-Davies solution of the Schr\"odinger equation here by solving the associated functions $\Gamma_n$ perturbatively to a given order in $\lambda$.  Fortunately, all that is needed to derive the part of the quantum Fokker-Planck equation that produces the standard stochastic Fokker-Planck equation is to compute these solutions to linear order in $\lambda$.  In fact, the zeroth order solution --- what would exist in the purely quadratic theory --- is already enough to find the stochastic noise.  By evaluating the order $\lambda$ parts of the solution as well, we shall obtain the correct drift term.  The advantage of this approach is that it is possible to generalise beyond the standard Fokker-Planck equation by simply working to higher orders in $\lambda$.

Let us begin by expanding each of the functions in the wave-functional as a power series in $\lambda$, 
\begin{eqnarray}
\alpha_k(t) &=& \sum_{n=0}^\infty \alpha_k^{(n)}(t) 
\nonumber \\
\Gamma_4(t;\vec k_1,\vec k_2,\vec k_3,\vec k_4) &=& 
\sum_{n=1}^\infty \Gamma_4^{(n)}(t;\vec k_1,\vec k_2,\vec k_3,\vec k_4)
\nonumber \\
\Gamma_6(t;\vec k_1,\vec k_2,\vec k_3,\vec k_4,\vec k_5,\vec k_6) 
&=& 
\sum_{n=2}^\infty \Gamma_6^{(n)}(t;\vec k_1,\vec k_2,\vec k_3,\vec k_4,\vec k_5,\vec k_6) ,
%\nonumber 
\end{eqnarray}
and so on.  The order in $\lambda$ is indicated by the corresponding superscript,
\begin{equation}
\alpha_k^{(n)}, \Gamma_4^{(n)}, \Gamma_6^{(n)}, \ldots \propto \lambda^n . 
\end{equation}
The higher order functions $\Gamma_n$ only begin their power series at correspondingly higher order in $\lambda$.  Because the trivial, Gaussian version of the theory already exists in the absence of any interactions, the leading term in the expansion of $\alpha_k(t)$ starts at zeroth order.  Exactly the same reasoning, tells us that $\Gamma_4$ and all of the higher order functions must vanish as $\lambda\to 0$.  In the quartic theory that we are analysing, $\Gamma_4$ itself starts with a linear term in the coupling $\lambda$, as is seen directly from its equation of motion.  But the equation for $\Gamma_6$, which we have not written explicitly here, is quadratic in $\Gamma_4$, so the series expansion for $\Gamma_6$ only begins with the $\lambda^2$ order term.  The problem of solving the Schr\"odinger equation to linear order in $\lambda$ then reduces to the problem of solving just three functions:  the zeroth and first order pieces of $\alpha_k(t)$, which we rename as $\bar\alpha_k(t)\equiv\alpha_k^{(0)}(t)$ and $\beta_k(t)\equiv\alpha_k^{(1)}(t)$ to avoid an excessive use of superscripts, and the leading part of $\Gamma_4$, 
\begin{eqnarray}
\alpha_k(t) &=& 
\bar\alpha_k(t) + \beta_k(t) + {\cal O}(\lambda^2)
\nonumber \\
\Gamma_4(t;\vec k_1,\vec k_2,\vec k_3,\vec k_4) &=& 
\Gamma_4^{(1)}(t;\vec k_1,\vec k_2,\vec k_3,\vec k_4) + {\cal O}(\lambda^2) .
%\nonumber 
\end{eqnarray}

The starting point is the purely Gaussian, noninteracting, theory, which is summarised by the function $\bar\alpha_k(t)$.  Even when $\alpha_k(t)$ has been shorn of its order $\lambda$ and higher parts, the differential equation for $\bar\alpha_k(t)$ is still nonlinear; so it is convenient to replace it with another function, $u_k(t)$ defined through 
\begin{equation}
\bar\alpha_k(t) = -i {\dot u_k(t)\over u_k(t)} . 
\end{equation}
The nonlinear, first-order equation for $\bar\alpha_k(t)$ then becomes a linear, second-order equation for $u_k(t)$, 
\begin{equation}
\ddot u_k + 3{\dot a\over a} \dot u_k + \biggl( m^2 + {k^2\over a^2} \biggr) u_k
= 0 .
\end{equation}
The vacuum, or Bunch-Davies, solution to this equation has the standard form, which is expressed more simply in terms of the conformal time coordinate $\eta$,
\begin{equation}
u_k(\eta) = {H\sqrt{\pi}\over 2} \eta^{3/2} H_\nu^{(2)}(k\eta) 
\qquad\hbox{where}\quad 
\nu^2 = {9\over 4} - {m^2\over H^2} ,
\end{equation}
and where $H_\nu^{(2)}(k\eta)$ is a Hankel function.  We have fixed this solution through two conditions.  One is the requirement that the solution should reproduce the vacuum solution in Minkowski space at scales $k\gg aH$, or in the limit $k\eta\to-\infty$.  Note that this condition fixes the single constant of integration that is needed to specify a particular solution for $\bar\alpha_k(t)$.  The second condition, which was not necessary for $\bar\alpha_k(t)$ but which has been used to normalise the function $u_k(t)$, is that we have required it to satisfy the condition, 
\begin{equation}
a^3\Bigl( u_k^*\dot u_k - u_k\dot u_k^*\Bigr) = -i .
\end{equation}
In the more frequently used Heisenberg picture for a free scalar field theory, this condition naturally emerges as the consequence of the equal-time commutation relation between the field and its conjugate momentum.  But in the Schr\"odinger picture, the overall normalisation always cancels within the ratio $\bar\alpha_k(t) = -i\dot u_k(t)/u_k(t)$.  Nonetheless, since the function $u_k(t)$ assumes a more recognisable form when we do impose this condition, we have chosen to use it here.  Taking the massless limit, the function $u_k(t)$ reduces to 
\begin{equation}
u_k(\eta) = {iH\over\sqrt{2}k^{3/2}} \bigl( 1+ik\eta\bigr) e^{-ik\eta} .
\end{equation}
If we then proceed to take the $k\eta\to-\infty$ limit too, we verify that the product $a(\eta)u_k(\eta)$ assumes the form of a Minkowski space vacuum mode for a massless theory,
$$
\lim_{k\eta\to-\infty} a(\eta)u_k(\eta) = {e^{-ik\eta}\over\sqrt{2k}} .
$$
From the perspective of a free, massless theory in Minkowski space, only the positive energies appear in the exponent; the negative energy solutions, $e^{ik\eta}$, are absent from this limiting form for $a(\eta)u_k(\eta)$.

Once we have found the leading part of the quadratic function $\alpha_k(t)$, we next compute the leading part of the function accompanying the quartic part of $\Gamma[\phi]$.  The series expansion for $\Gamma_6$ only begins at quadratic order, so we shall not need to include this function when solving for just the leading part of $\Gamma_4^{(1)}$.  Without the $\Gamma_6$ term, the linear part of the equation for $\Gamma_4$ in $\lambda$ reduces to a first-order inhomogeneous equation,
\begin{equation}
{\partial\Gamma_4^{(1)}\over\partial t} 
+ {\partial\over\partial t}\biggl[ 
\ln \Bigl( a^3 u_{k_1} u_{k_2} u_{k_3} u_{k_4} \Bigr) \biggr]
\Gamma_4^{(1)} 
= i\lambda .
\end{equation}
Its general solution is 
\begin{equation}
\Gamma_4^{(1)}(\eta;\vec k_1,\vec k_2,\vec k_3,\vec k_4)
= {\displaystyle c_4 + i\lambda \int^\eta_{\eta_0} d\eta' a^4(\eta') 
u_{k_1}(\eta') u_{k_2}(\eta') u_{k_3}(\eta') u_{k_4}(\eta')\over 
a^3(\eta) u_{k_1}(\eta) u_{k_2}(\eta) u_{k_3}(\eta) u_{k_4}(\eta)} ,
\end{equation}
where the constant of integration, $c_4$, should be fixed by the requirement that the wave-functional corresponds to the Bunch-Davies state.  To find the correct choice for this constant, we again consider the behaviour of $\Gamma_4$ when all of the wavelengths are much smaller that the size of the horizon.  In this limit, where $k_i\eta\to -\infty$, we can replace  
$$
a(\eta)u_{k_i}(\eta) \approx {e^{-ik_i\eta}\over\sqrt{2k_i}} 
$$
in the solution for $\Gamma_4^{(1)}(\eta;\vec k_1,\vec k_2,\vec k_3,\vec k_4)$.  Being able unambiguously to perform the integral that appears in the solution depends on being able to establish a suitable $i\epsilon$ prescription for the time-integration contour.  As one proceeds ever deeper into the horizon by allowing the initial time to reach further back, we should define this $i\epsilon$ prescription so that the initial contribution to the integral in the solution for $\Gamma_4^{(1)}$ vanishes as $\eta_0\to-\infty$.  Once this has been done, if we consider times where the momenta are still well within the horizon at $\eta$ --- that is, $-k_i\eta\gg 1$ --- the leading behaviour that results when performing the integral is 
\begin{equation}
\Gamma_4^{(1)}(\eta;\vec k_1,\vec k_2,\vec k_3,\vec k_4)
\approx c_4\, {a(\eta)\sqrt{16k_1k_2k_3k_4}\over e^{-i(k_1+k_2+k_3+k_4)\eta}}
- {a(\eta)\, \lambda\over k_1+k_2+k_3+k_4} .
\end{equation}
We now realise that the $i\epsilon$ prescription that has successfully suppressed the unwanted contribution from the positive energy fluctuations as $\eta_0\to -\infty$ would correspondingly lead to an exponential growth of the first term as $k_i\eta\to-\infty$.  By choosing $c_4=0$, this problem is resolved since the first term has been removed entirely, along with what would appear to be negative energy oscillations from the perspective of an observer only able to measure wavelengths much smaller than the size of the horizon.   Thus, the leading part of the quartic function for the Bunch-Davies state is 
\begin{equation}
\Gamma_4^{(1)}(\eta;\vec k_1,\vec k_2,\vec k_3,\vec k_4)
= {i\lambda \displaystyle\int^\eta_{-\infty} d\eta' a^4(\eta') 
u_{k_1}(\eta') u_{k_2}(\eta') u_{k_3}(\eta') u_{k_4}(\eta') \over 
a^3(\eta) u_{k_1}(\eta) u_{k_2}(\eta) u_{k_3}(\eta) u_{k_4}(\eta)} .
\end{equation}
 
Now that we have found the appropriate solution for our state, we can investigate how it behaves in the opposite limit --- it is the set of long wavelength fluctuations that are relevant for the stochastic description of the theory.  For a massless field, the integral is once again simple enough to evaluate explicitly,
\begin{eqnarray}
\Gamma_4^{(1)}(\eta;\vec k_1,\vec k_2,\vec k_3,\vec k_4)
&=&
{i\lambda\over 3H} {\displaystyle 1 + iK\eta - {1\over 2}K^2\eta^2 
+ {3\over 2}\bigl( k_1^2+k_2^2+k_3^2+k_4^2 \bigr) \eta^2
- {3ik_1k_2k_3k_4\eta^3\over K}
\over (1+ik_1\eta)(1+ik_2\eta)(1+ik_3\eta)(1+ik_4\eta)}
\nonumber \\
&&
+ {\lambda\over 3H} {\bigl( k_1^3+k_2^3+k_3^3+k_4^3 \bigr)\eta^3 e^{iK\eta}\, 
{\rm Ei}(1, iK\eta)\over (1+ik_1\eta)(1+ik_2\eta)(1+ik_3\eta)(1+ik_4\eta)} .
%\nonumber 
\end{eqnarray}
Here we have abbreviated $K\equiv k_1+k_2+k_3+k_4$ and ${\rm Ei}(1,iK\eta)$ is the standard exponential integral function.  The advantage of analysing the theory in the Schr\"odinger picture is becoming more apparent --- this function, which is the one accompanying the quartic term in $\Gamma[\phi]$, is completely free from any divergent behaviour in the long wavelength limit where $k_i\eta\to 0$.  The exponential integral diverges logarithmically when its argument approaches zero,
$$
{\rm Ei}(1, iK\eta) = - \gamma + \ln(iK\eta) + iK\eta + {\cal O}(K^2\eta^2) ,
$$
but since this only happens when all four of the momenta simultaneously become small, and since the exponential integral is multiplied by $( k_1^3+k_2^3+k_3^3+k_4^3)\eta^3$, there are no long wavelength divergences in $\Gamma_4^{(1)}$.

Later, we shall see that the asymptotic behaviour of this function fixes the drift in the quantum Fokker-Planck equation and --- because they are precisely the same --- the drift in the stochastic Fokker-Planck equation as well.  This function is genuinely produced by the interactions amongst the fields, so it is directly responsible for the existence of the terms in the Fokker-Planck equation that are associated with the potential.  For this reason, we need to evaluate $\Gamma_4^{(1)}$ in the limit where all of its momenta have been stretched far outside the horizon, $k_i<\varepsilon aH$.  In this case, the function approaches a purely imaginary constant, 
\begin{equation}
\lim_{{k_i\eta\to 0\atop m=0}} 
\Gamma_4^{(1)}(\eta;\vec k_1,\vec k_2,\vec k_3,\vec k_4)
= {i\over 3} {\lambda\over H} + {i\over 2} {\lambda\over H} \bigl( k_1^2+k_2^2+k_3^2+k_4^2 \bigr)\eta^2 + \cdots .
\end{equation}
In the opposite limit, where the wavelengths of the fluctuations labelled by $\vec k_i$ are all well within the horizon, $k_i\gg aH$, we see that this function is more and more suppressed, 
\begin{equation}
\lim_{{k_i\eta\to-\infty\atop m=0}} 
\Gamma_4^{(1)}(\eta;\vec k_1,\vec k_2,\vec k_3,\vec k_4)
= {\lambda\over H} {1\over (k_1+k_2+k_3+k_4)\eta} + \cdots .
\end{equation}
So in a sense, at early times and for short wavelengths the theory assumes a more and more strongly Gaussian character.

Although the quadratic part of the wave-functional $\alpha_k(t)$ also contains parts that scale as $\lambda$, which are the same order in the coupling as the leading behaviour of the quartic function $\Gamma_4(\eta;\vec k_1,\vec k_2,\vec k_3,\vec k_4)$, it turns out that for the purpose of treating the static solutions of the Fokker-Planck equation it would not be consistent to include these order $\lambda$ terms in $\alpha_k$ without having also included the order $\lambda^2$ terms of $\Gamma_4$ as well.  To understand why this is so, let us consider the static limit of the stochastic theory where $\partial p/\partial t=0$.  The Fokker-Planck equation then reduces to the equation 
\begin{equation}
N {\partial^2p\over\partial\varphi^2} 
+ D {\partial\over\partial\varphi}\biggl( 
{\partial V\over\partial\varphi} p(\varphi) \biggr) = 0 ,
\end{equation}
whose general solution is 
\begin{equation}
p(\varphi) = n \biggl[ e^{-{D\over N}V(\varphi)}
+ c e^{-{D\over N}V(\varphi)} \int^\varphi d\varphi'\, e^{{D\over N}V(\varphi')} 
\biggr] , 
\end{equation}
where $n$ is the normalisation of the probability function.  Choosing $c=0$, the static solution for a quartic interaction is 
\begin{equation}
p(\varphi) = n e^{-{D\over N}V(\varphi)}
= {\Gamma\bigl({3\over 4}\bigr)\over\pi} 
\biggl( {\lambda D\over 6N} \biggr)^{1/4} e^{-{\lambda D\over 24N}\varphi^4} . 
\end{equation}
The effect of an order $\lambda$ term in $N$ would only produce an order $\lambda^2$ effect in the ratio $\lambda D/N$.  This would be exactly the same order as the next term in the series expansion of $\Gamma_4$ contributing to the drift.

It is nonetheless useful at this stage to investigate a little of the asymptotic behaviour of $\beta_k(t)$  --- the order $\lambda$ part of the quadratic function --- at late times,
$$
\alpha_k(t) = \bar\alpha_k(t) + \beta_k(t) + \cdots 
= - i{\dot u_k(t)\over u_k(t)} + \beta_k(t) + {\cal O}(\lambda^2) . 
$$
By extracting the order $\lambda$ terms from the differential equation for $\dot\alpha_k(t)$, we obtain a differential equation for $\beta_k(t)$,
\begin{equation}
{\partial\over\partial t}\bigl( u_k^2 a^3\beta_k \bigr) 
= i u_k^2 \biggl[ a^3\delta m^2 + (Z_1-1)ak^2 
+ {1\over 2} \int {d^3\vec p\over (2\pi)^3}\, \Gamma_4^{(1)}(t;\vec k,-\vec k,\vec p,-\vec p) \biggr] .
\end{equation}
Since the integral of the quartic function diverges at short wavelengths, we must introduce counterterms, $\delta m^2$ and $(Z_1-1)$, which were not necessary for computing the leading behaviour of either $\alpha_k(t)$ or $\Gamma_4$.  In a massless theory, we already showed that $\Gamma_4^{(1)}$ approaches an imaginary constant for long wavelengths, so there are no divergences in this integral associated with small values of $p=|\!|\vec p|\!|$.  The general solution of this equation is 
$$
\beta_k(t) = {1\over a^3(t) }{i\over u_k^2(t)}\biggl\{ c_2 
+ \int^t dt'\, u_k^2(t') \biggl[ a^3(t')\delta m^2 + (Z_1-1)a(t')k^2 
+ {1\over 2} \int {d^3\vec p\over (2\pi)^3}\, \Gamma_4^{(1)}(t';\vec k,-\vec k,\vec p,-\vec p) \biggr] \biggr\} .
$$
Once again the $i\epsilon$ prescription for the Bunch-Davies state, which suppresses the contribution from the lower end of the time integral associated with fluctuations that are infinitesimally tiny when compared with the size of the horizon, would also cause the term proportional to $c_2$ to diverge at early times.  The appropriate choice for this state is $c_2=0$.  The order $\lambda$ part of the quadratic function for the Bunch-Davies state is then
\begin{equation}
\beta_k(\eta) = {1\over a^3(\eta) }{i\over u_k^2(\eta)} 
\int_{-\infty}^\eta d\eta'\, u_k^2(\eta') \biggl[ 
a^4(\eta')\delta m^2 + (Z_1-1)a^2(\eta')k^2 
+ {1\over 2} a(\eta')\int {d^3\vec p\over (2\pi)^3}\, \Gamma_4^{(1)}(\eta';\vec k,-\vec k,\vec p,-\vec p) \biggr] ,
\end{equation}
when expressed as a function of the conformal time coordinate.

We now analyse the behaviour of $\beta_k(t)$ at long wavelengths to see that it is well behaved.  Looking at the explicit form for $u_k(\eta)$ for a massless theory, it is clear that as long as the result of performing the momentum integral in this solution does not diverge faster than $1/\eta^3$, $\beta_k(\eta)$ will not itself diverge as $k\eta\to 0$.  By rescaling the momentum $\vec p$ by $\vec p\eta'$, we can express the momentum integral in this solution in terms of a dimensionless function,
\begin{equation}
\int {d^3\vec p\over (2\pi)^3}\, \Gamma_4^{(1)}(\eta';\vec k,-\vec k,\vec p,-\vec p) 
= {1\over\eta^{\prime 3}}{\lambda\over 3H} \int {d^3(\vec p\eta')\over (2\pi)^3}\, \hat\Gamma_4^{(1)}(k\eta',p\eta') ,
\end{equation}
where
\begin{eqnarray}
\hat\Gamma_4^{(1)}(k\eta,p\eta)
&=& {\displaystyle i - 2(p+k)\eta + i(p^2-4pk+k^2)\eta^2 
+ {3\over 2}{k^2p^2\eta^3\over (p+k)} \over (1+ik\eta)^2(1+ip\eta)^2}
\nonumber \\
&& 
+ {2\bigl( p^3+k^3 \bigr)\eta^3 e^{2i(p+k)\eta}\, 
{\rm Ei}(1, 2i(p+k)\eta) \over (1+ik\eta)^2(1+ip\eta)^2}
%\nonumber 
\end{eqnarray}
is the dimensionless quantity.  Since $\Gamma_4^{(1)}$ behaves well at long wavelengths, it cannot diverge as $k\eta'\to 0$ which in turn means that its integral cannot diverge faster than the $1/\eta^{\prime 3}$ factor that we have already extracted.  In fact, the leading $1/\eta^{\prime 3}$ scaling has the same power as a mass term.  For a massless theory, this term should be absent, as can be arranged --- depending on the regularization scheme being used --- through a suitable choice for $\delta m^2$.  Therefore, in a massless theory $\beta_k(\eta)$ should vanish as $k\eta\to 0$.

\section{The quantum Fokker-Planck equation\/} 

\noindent 
We are ready to use what we have learned to derive a quantum version of the Fokker-Planck equation.  To do so, we must solve for the evolution of the diagonal part of the density matrix for the Bunch-Davies state, $P[\phi]=\Psi[\phi]\Psi^*[\phi]$, and from it derive the evolution of the density matrix for the coarsely grained version of the theory, $P_\Omega[\phi_L]$.  The latter is the density matrix obtained by integrating out the short wavelength fluctuations, 
\begin{equation}
P_\Omega[\phi_L] 
\equiv \int_S {\cal D}\phi_{\vec p}\, P[\phi] 
= \int_{p\ge \varepsilon aH}\hskip-12truept {\cal D}\phi_{\vec p}\, P[\phi] .
\end{equation}
The time derivative of $P_\Omega[\phi_L]$ will then produce the quantum version of the Fokker-Planck equation that we are seeking.  One subtlety that occurs in this effective theory, and which does not usually happen in most standard effective field theories, is that in taking the time derivative of $P_\Omega[\phi_L]$ we must also include the time dependence that occurs in the boundary, $\varepsilon aH$, dividing the long wavelength fluctuations that we must keep from the short wavelength ones that we remove.

The important idea here is to match between the two theories.  This step allows us to express the functions inside the density matrix of the effective theory in terms of those of the original theory.  We can then use the Schr\"odinger equation of the full theory to compute the time derivative of $P_\Omega[\phi_L]$ directly.  In terms of the wave-functional $\Psi[\phi]$ for the Bunch-Davies state, the diagonal part of its density matrix is 
\begin{equation}
P[\phi] = \Psi[\phi]\Psi^*[\phi] 
= |N|^2 e^{-a^3[\Gamma[\phi]+\Gamma^*[\phi]]} .
\end{equation}
The first step in the matching process is to define an analogous expansion for the density matrix of the effective theory,
\begin{equation}
P_\Omega[\phi_L] = |N_\Omega|^2 e^{-a^3[\Gamma_\Omega[\phi_L]+\Gamma_\Omega^*[\phi_L]]} ,
\end{equation}
where
\begin{equation}
\Gamma_\Omega[\phi_L] = \sum_{n=2}^\infty {1\over n!} 
\int_L {d^3k_1\over (2\pi)^3} \cdots {d^3k_n\over (2\pi)^3}\, 
(2\pi)^3\delta^3(\vec k_1+\cdots +\vec k_n)\,
\Gamma_{\Omega,n}(t;\vec k_1,\ldots,\vec k_n)\, \phi_{\vec k_1} 
\cdots \phi_{\vec k_n} ,
\end{equation}
together with its own normalisation $N_\Omega$ which satisfies the condition
\begin{equation}
\int_L {\cal D}\phi_{\vec k}\, P[\phi_L] = 1 .
\end{equation}

In order to calculate the density matrix of the effective theory, it is helpful to separate the full density matrix into three factors,
\begin{equation}
P[\phi] = |N|^2 e^{-a^3[\Gamma+\Gamma^*]} 
= |N|^2 e^{-a^3[\Gamma_L+\Gamma_L^*]} e^{-a^3[\Gamma_0+\Gamma_0^*]} 
e^{-a^3[\delta\Gamma_S+\delta\Gamma_S^*]} ,
\end{equation}
which we have expressed through an equivalent separation of $\Gamma[\phi_L,\phi_S]$ into three terms:  one that only includes the long wavelength modes, 
\begin{equation}
\Gamma_L[\phi_L] 
= \sum_{n=2}^\infty {1\over n!} \int_L {d^3\vec k_1\over (2\pi)^3} \cdots 
{d^3\vec k_n\over (2\pi)^3}\, (2\pi)^3 \delta^3(\vec k_1 + \cdots + \vec k_n)\, 
\phi_{\vec k_1}\cdots\phi_{\vec k_n} \Gamma_n(\vec k_1, \ldots, \vec k_n) 
\end{equation}
one that is quadratic in the short wavelength modes and zeroth order in $\lambda$,
\begin{equation}
\Gamma_0[\phi_S] = {1\over 2} \int_S {d^3\vec p\over (2\pi)^3}\, 
\phi_{\vec p}\phi_{-\vec p} \bar\alpha_p ,
\end{equation}
and a final term that collects everything else,
\begin{eqnarray}
\delta\Gamma_S[\phi_L,\phi_S] &=& 
{1\over 2} \int_S {d^3\vec p\over (2\pi)^3}\, 
\phi_{\vec p}\phi_{-\vec p} \bigl( \alpha_p - \bar\alpha_p \bigr)
 \\
&& 
+ \sum_{n=4}^\infty {1\over n!} \int_{\not\hskip-0.75truept L} {d^3\vec p_1\over (2\pi)^3} \cdots 
{d^3\vec p_n\over (2\pi)^3}\, (2\pi)^3 \delta^3(\vec p_1 + \cdots + \vec p_n)\, 
\phi_{\vec p_1}\cdots\phi_{\vec p_n} \Gamma_n(\vec p_1, \ldots, \vec p_n) . 
\nonumber 
\end{eqnarray}
Here the notation $\not\hskip-4truept L$ means that at least one of the momentum integrals is over just the short wavelength modes; integrals over long wavelengths can appear in this term too.

Integrating out the short wavelength fluctuations of the fields is done the most straightforwardly by further separating the purely noninteracting part of the density matrix for the short-distance degrees of freedom from the rest, defining in the process 
\begin{equation}
P_0[\phi_S] \equiv |N_0| \exp\biggl\{ - {1\over 2} a^3 \int_S {d^3\vec p\over (2\pi)^3}\, \phi_{\vec p}\phi_{-\vec p} \bigl( \bar\alpha_p + \bar\alpha_p^* \bigr) 
\biggr\} ,
\qquad
\int_S {\cal D}\phi_{\vec p}\, P_0[\phi_S] = 1 ,
\end{equation}
and working perturbatively in the coupling.  Since we are only evaluating the coarse-grained density matrix $P_\Omega[\phi_L]$ to linear order in $\lambda$, we can afford to be a little sloppy and move the effect of integrating the $\delta\Gamma_S$ term directly into the exponent,
\begin{eqnarray}
P_\Omega[\phi_L] &=& 
|N|^2 e^{-a^3[\Gamma_L+\Gamma_L^*]} \int_S {\cal D}\phi_{\vec p}\, 
e^{-a^3[\Gamma_0+\Gamma_0^*]} e^{-a^3[\delta\Gamma_S+\delta\Gamma_S^*]}
\nonumber \\
&=& 
{|N|^2\over |N_0|^2} e^{-a^3[\Gamma_L+\Gamma_L^*]} \int_S {\cal D}\phi_{\vec p}\, 
P_0[\phi_S] \Bigl[ 1 -a^3[\delta\Gamma_S+\delta\Gamma_S^*] + \cdots \Bigr]
\nonumber \\
&=& 
{|N|^2\over |N_0|^2} e^{-a^3[\Gamma_L+\Gamma_L^*]} \biggl[ 
1 - a^3 \int_S {\cal D}\phi_{\vec p}\, 
P_0 [\delta\Gamma_S+\delta\Gamma_S^*] + \cdots
\biggr]
\nonumber \\
&=& 
{|N|^2\over |N_0|^2} e^{-a^3[\Gamma_L+\Gamma_L^*]
- a^3 \int_S {\cal D}\phi_{\vec p}\, 
P_0 [\delta\Gamma_S+\delta\Gamma_S^*] } + {\cal O}(\lambda^2).  
\end{eqnarray}
The exponent is not quite yet meant to be identified with $-a^3[\Gamma_\Omega +\Gamma_\Omega^*]$, as it also contains contributions to the normalisation of the coarse-grained density matrix.  These contributions are easily recognised since they do not contain any factors of the field, $\phi_{\vec k}$, and they are accompanied by the usual infinite factor $(2\pi)^3\, \delta^3(\vec 0)$ associated with the infinite spatial volume.  When we perform\footnote{We only require terms up to the quadratic order for which the following integrals are sufficient,
\begin{eqnarray}
\int_S {\cal D}\phi_{\vec p}\, \phi_{\vec p_1}\phi_{\vec p_2} P_0[\phi_S] 
&=& 
{1\over a^3} {1\over\bar\alpha_{p_1}+\bar\alpha_{p_1}^*} 
(2\pi)^3\delta^3(\vec p_1+\vec p_2)
\nonumber \\
\int_S {\cal D}\phi_{\vec p}\, 
\phi_{\vec p_1}\phi_{\vec p_2}\phi_{\vec p_3}\phi_{\vec p_4} P_0[\phi_S] 
&=& 
{1\over a^6} 
{(2\pi)^3\delta^3(\vec p_1+\vec p_2)\over\bar\alpha_{p_1}+\bar\alpha_{p_1}^*} 
{(2\pi)^3\delta^3(\vec p_3+\vec p_4)\over\bar\alpha_{p_3}+\bar\alpha_{p_3}^*} 
+ {1\over a^6} 
{(2\pi)^3\delta^3(\vec p_1+\vec p_3)\over\bar\alpha_{p_1}+\bar\alpha_{p_1}^*} 
{(2\pi)^3\delta^3(\vec p_2+\vec p_4)\over\bar\alpha_{p_2}+\bar\alpha_{p_2}^*} 
\nonumber \\
&&
+ {1\over a^6} 
{(2\pi)^3\delta^3(\vec p_1+\vec p_4)\over\bar\alpha_{p_1}+\bar\alpha_{p_1}^*} 
{(2\pi)^3\delta^3(\vec p_2+\vec p_3)\over\bar\alpha_{p_2}+\bar\alpha_{p_2}^*} .
\nonumber 
\end{eqnarray}}
the functional integrals that occur in the exponent, we obtain
\begin{eqnarray}
&&\hskip-0.25truein
\Gamma_\Omega[\phi_L] - a^{-3} \delta N
\nonumber \\
&=&
\Gamma_L[\phi_L] 
+ \int_S {\cal D}\phi_{\vec p}\, P_0[\phi_S]\, \delta\Gamma_S[\phi_L,\phi_S]
+ {\cal O}(\lambda^2)
\nonumber \\
&=&
{1\over 2} \int_L {d^3\vec k\over (2\pi)^3}\, 
\phi_{\vec k}\phi_{-\vec k} \biggl[ 
\alpha_k 
+ {1\over 2} {1\over a^3} 
\int_S {d^3\vec p\over (2\pi)^3}\, {1\over\bar\alpha_p+\bar\alpha_p^*}
\Gamma_4^{(1)}(\vec k, -\vec k, \vec p, -\vec p) 
\biggr]
\nonumber \\
&&
+ {1\over 4!} \int_L {d^3\vec k_1\over (2\pi)^3}{d^3\vec k_2\over (2\pi)^3}
{d^3\vec k_3\over (2\pi)^3}{d^3\vec k_4\over (2\pi)^3}\, 
(2\pi)^3 \delta^3(\vec k_1 + \vec k_2 + \vec k_3 + \vec k_4)\, 
\phi_{\vec k_1}\phi_{\vec k_2}\phi_{\vec k_3}\phi_{\vec k_4} 
\Gamma_4^{(1)}(\vec k_1, \vec k_2, \vec k_3, \vec k_4) 
\nonumber \\
&&
+ {1\over 2} {1\over a^3} (2\pi)^3 \delta^3(\vec 0) 
\int_S {d^3\vec p\over (2\pi)^3}\, 
{1\over\bar\alpha_p+\bar\alpha_p^*} 
\biggl[ \beta_p
+ {1\over 4} {1\over a^3} \int_S {d^3\vec p'\over (2\pi)^3}\, 
{1\over\bar\alpha_{p'}+\bar\alpha_{p'}^*}
\Gamma_4^{(1)}(\vec p, -\vec p, \vec p', -\vec p') 
\biggr] 
\nonumber \\
&&
+ {\cal O}(\lambda^2) . 
\end{eqnarray}
Matching the terms with two, four, or no factors of the field $\phi_{\vec k}$ produces the following functions that describe the evolution of the coarse-grained density matrix,
\begin{eqnarray}
\alpha_{\Omega,k} \equiv \Gamma_{\Omega,2}(t;\vec k,-\vec k)
&=&
\bar\alpha_k + \beta_k 
+ {1\over 2} {1\over a^3} \int_S {d^3\vec p\over (2\pi)^3}\, 
{1\over\bar\alpha_p+\bar\alpha_p^*}
\Gamma_4^{(1)}(t;\vec k, -\vec k, \vec p, -\vec p) 
+ {\cal O}(\lambda^2) 
\nonumber \\
\Gamma_{\Omega,4}(t;\vec k_1, \vec k_2, \vec k_3, \vec k_4) 
&=&
\Gamma_4^{(1)}(t;\vec k_1, \vec k_2, \vec k_3, \vec k_4) 
+ {\cal O}(\lambda^2) ,  
\end{eqnarray}
and
\begin{equation}
\delta N = - {1\over 2} (2\pi)^3 \delta^3(\vec 0) 
\int_S {d^3\vec p\over (2\pi)^3}\, 
{1\over\bar\alpha_p+\bar\alpha_p^*} 
\biggl[ \beta_p
+ {1\over 4} {1\over a^3} \int_S {d^3\vec p'\over (2\pi)^3}\, 
{1\over\bar\alpha_{p'}+\bar\alpha_{p'}^*}
\Gamma_4^{(1)}(\vec p, -\vec p, \vec p', -\vec p') 
\biggr] 
+ {\cal O}(\lambda^2) ,
\end{equation}
where the normalisation of the density matrix of the effective theory is 
\begin{equation}
|N_\Omega|^2 = {|N|^2\over |N_0|^2} e^{\delta N+\delta N^*} . 
\end{equation}

The density matrix of the coarsely grained theory now inherits its time dependence directly from the original theory.  For example, the time derivative of the leading parts of $\alpha_{\Omega,k}$ follows from how $\alpha_k(t)$ and $\Gamma_4^{(1)}$ evolve, 
\begin{eqnarray}
{\partial\alpha_{\Omega,k}\over\partial t} + 3{\dot a\over a} \alpha_{\Omega,k} 
&=& 
- i\alpha_{\Omega,k}^2
+ {i\over 2}{1\over a^3} \int_L {d^3\vec k'\over (2\pi)^3}\, 
\Gamma_4^{(1)}(t;\vec k,-\vec k,\vec k',-\vec k')
\nonumber \\
&& 
+ {1\over 2} {1\over a^3} \int_{\partial S} {d^3\vec p\over (2\pi)^3}\, 
{1\over\bar\alpha_p+\bar\alpha_p^*}
\Gamma_4^{(1)}(t;\vec k, -\vec k, \vec p, -\vec p) 
\nonumber \\
&& 
+ i\biggl\{ m^2 + {k^2\over a^2} 
+ {1\over 2}{\lambda\over a^3} \int_S {d^3\vec p\over (2\pi)^3}\, 
{1\over\bar\alpha_p+\bar\alpha_p^*}
\biggr\} 
+ \cdots . 
\end{eqnarray}
The terms on the first line are all expressed in terms of the functions of the effective theory.  Those appearing on the third line are purely imaginary and cancel within the combinations of $\alpha_{\Omega,k}(t)+\alpha_{\Omega,k}^*(t)$ that occur in the density matrix.  The only unfamiliar term is the one on the second line.  It arises because when we truncate the momenta, $k\le \varepsilon a(t)H$, the limit of the truncated integral is also time dependent.  Introducing this boundary as a step function, its time derivative only contributes at the boundary.  This has been denoted about with the following notation,
\begin{eqnarray}
\int_{\partial L} {d^3\vec k\over (2\pi)^3}\, f(\vec k)
&\equiv& 
\int {d^3\vec k\over (2\pi)^3}\, f(\vec k) 
{\partial\over\partial t} \Theta(\varepsilon aH-k)
\nonumber \\
\int_{\partial S} {d^3\vec k\over (2\pi)^3}\, f(\vec k)
&\equiv& 
\int {d^3\vec k\over (2\pi)^3}\, f(\vec k) 
{\partial\over\partial t} \Theta(k-\varepsilon aH) ,
\nonumber 
\end{eqnarray}
where $f(\vec k)$ is a general function of the momentum.

Combining the appropriate derivative and its complex conjugate, the evolution of the quadratic structure in $P_\Omega[\phi_L]$ is summarised by 
\begin{eqnarray}
&&\hskip-0.375truein
{\partial\over\partial t} \Bigl[ a^3\, \bigl( \alpha_{\Omega,k} + \alpha_{\Omega,k}^* \bigr) \Bigr] 
\nonumber \\
&=& 
- ia^3\bigl( \alpha_{\Omega,k}^2 - \alpha_{\Omega,k}^{*2} \bigr)
+ {i\over 2} \int_L {d^3\vec k'\over (2\pi)^3}\, 
\bigl[ \Gamma_4^{(1)}(t;\vec k,-\vec k,\vec k',-\vec k')
- \Gamma_4^{(1)*}(t;-\vec k,\vec k,-\vec k',\vec k')
\bigr] 
\nonumber \\
&& 
+ {1\over 2} {1\over a^3} \int_{\partial S} {d^3\vec p\over (2\pi)^3}\, 
{1\over\bar\alpha_p+\bar\alpha_p^*}
\bigl[ \Gamma_4^{(1)}(t;\vec k, -\vec k, \vec p, -\vec p) 
+ \Gamma_4^{(1)*}(t;\vec k, -\vec k, \vec p, -\vec p) \bigr] 
+ \cdots 
\end{eqnarray}
to linear order in the coupling.  At this order the evolution of the quartic term in the fields is precisely the same as in the original theory.  Finally, the evolution of the normalisation of the coarsely grained density matrix follows from
\begin{equation}
{\dot N_\Omega\over N_\Omega} + {\dot N_\Omega^*\over N_\Omega^*} 
= - {i\over 2} (2\pi)^3\delta^3(\vec 0) \int_L {d^3\vec k\over (2\pi)^3}\, 
\bigl( \alpha_{\Omega,k} - \alpha_{\Omega,k}^* \bigr)
+ {1\over 2} (2\pi)^3\delta^3(\vec 0) \int_{\partial L} {d^3\vec k\over (2\pi)^3}.
\end{equation}

We are now able to evaluate the time derivative of $P_\Omega[\phi_L]$ directly,
\begin{equation}
i{\partial P_\Omega\over\partial t} 
= \biggl\{ i{\dot N_\Omega\over N_\Omega} + i{\dot N_\Omega^*\over N_\Omega^*} 
- i {\partial\over\partial t} \Bigl[ a^3 \bigl( \Gamma_\Omega + \Gamma_\Omega^* \bigr) \Bigr] \biggr\} ,
\end{equation}
through our knowledge of how each of the functions associated with the original theory itself evolves.  When we do so, we obtain the following expression, 
\begin{eqnarray}
i{\partial P_\Omega\over\partial t} &=& 
\biggl\{ 
- {1\over 2} a^3 \int_L {d^3\vec k\over (2\pi)^3}\, \phi_{\vec k} \phi_{-\vec k} 
\bigl( \alpha_{\Omega,k}^2 - \alpha_{\Omega,k}^{*2} \bigr)
- {i\over 2} a^3 \int_{\partial L} {d^3\vec k\over (2\pi)^3}\, \phi_{\vec k} \phi_{-\vec k} 
\bigl( \alpha_{\Omega,k} + \alpha_{\Omega,k}^* \bigr)
\nonumber \\
&& 
+ {1\over 4} \int_L {d^3\vec k\over (2\pi)^3}\, \phi_{\vec k} \phi_{-\vec k} 
\int_L {d^3\vec k'\over (2\pi)^3}\, \Bigl[ 
\Gamma^{(1)}_4(\vec k, -\vec k, \vec k', -\vec k') 
- \Gamma^{(1)*}_4(-\vec k, \vec k, -\vec k', \vec k') 
\Bigr]
\nonumber \\
&& 
- {i\over 4} \int_L {d^3\vec k\over (2\pi)^3}\, \phi_{\vec k} \phi_{-\vec k} 
\int_{\partial S} {d^3\vec p\over (2\pi)^3}\, 
{1\over\bar\alpha_p + \bar\alpha_p^*} \Bigl[ 
\Gamma^{(1)}_4(\vec k, -\vec k, \vec p, -\vec p) 
+ \Gamma^{(1)*}_4(-\vec k, \vec k, -\vec p, \vec p) 
\Bigr]
\nonumber \\
&& 
- {1\over 4!} a^3  
\int_L {d^3\vec k_1\over (2\pi)^3} {d^3\vec k_2\over (2\pi)^3}
{d^3\vec k_3\over (2\pi)^3} {d^3\vec k_4\over (2\pi)^3}\, 
(2\pi)^3\, \delta^3(\vec k_1 + \vec k_2 + \vec k_3 + \vec k_4)\, 
\phi_{\vec k_1} \phi_{\vec k_2} \phi_{\vec k_3} \phi_{\vec k_4} 
\nonumber \\
&&\qquad\qquad \times 
\Bigl[ \bigl[ \bar\alpha_{k_1} + \bar\alpha_{k_2} 
+ \bar\alpha_{k_3} + \bar\alpha_{k_4} \bigr]
\Gamma^{(1)}_4(\vec k_1, \vec k_2, \vec k_3, \vec k_4)  
\nonumber \\
&&\qquad\qquad\quad 
- \bigl[ \bar\alpha^*_{k_1} + \bar\alpha^*_{k_2} 
+ \bar\alpha^*_{k_3} + \bar\alpha^*_{k_4} \bigr]
\Gamma^{(1)*}_4(-\vec k_1, -\vec k_2, -\vec k_3, -\vec k_4) 
\Bigr] 
\nonumber \\
&& 
- {i\over 4!} a^3  
\int_{\partial L} {d^3\vec k_1\over (2\pi)^3} {d^3\vec k_2\over (2\pi)^3}
{d^3\vec k_3\over (2\pi)^3} {d^3\vec k_4\over (2\pi)^3}\, 
(2\pi)^3\, \delta^3(\vec k_1 + \vec k_2 + \vec k_3 + \vec k_4)\, 
\phi_{\vec k_1} \phi_{\vec k_2} \phi_{\vec k_3} \phi_{\vec k_4} 
\nonumber \\
&&\qquad\qquad \times 
\Bigl[ \Gamma^{(1)}_4(\vec k_1, \vec k_2, \vec k_3, \vec k_4)  
+ \Gamma^{(1)*}_4(-\vec k_1, -\vec k_2, -\vec k_3, -\vec k_4) 
\Bigr] 
\nonumber \\
&& 
+ {1\over 2} (2\pi)^3\delta^3(\vec 0) \int_L {d^3\vec k\over (2\pi)^3}\, 
\bigl( \alpha_{\Omega,k} - \alpha_{\Omega,k}^* \bigr)
+ {i\over 2} (2\pi)^3\delta^3(\vec 0) \int_{\partial L} {d^3\vec k\over (2\pi)^3}
+ {\cal O}(\lambda^2) \biggr\} P_\Omega . 
\nonumber \\
\end{eqnarray}
This is essentially the Liouville equation for the effective theory.  It is also the quantum Fokker-Planck equation, as we shall now show.  The first step is to translate some of these terms into a second functional derivative of $P_\Omega[\phi_L]$ with the appropriate coefficient.  For this purpose, the following formula, written for an arbitrary momentum-dependent coefficient, $F_k$, is very useful,
\begin{eqnarray}
&&\hskip-0.30truein
{1\over a^3} \int_L {d^3\vec k\over (2\pi)^3}\, F_k
{\delta^2P_\Omega[\phi_L]\over\delta\phi_{\vec k}\delta\phi_{-\vec k}} 
\nonumber \\
&=& 
\biggl\{
- (2\pi)^3\, \delta^3(\vec 0)
\int_L {d^3\vec k\over (2\pi)^3}\, F_k 
\bigl[ \alpha_{\Omega,k}(t) + \alpha_{\Omega,k}^*(t) \bigr]
+ a^3 \int_L {d^3\vec k\over (2\pi)^3}\, \phi_{\vec k} \phi_{-\vec k}\, 
F_k \bigl[ \alpha_{\Omega,k}(t) + \alpha_{\Omega,k}^*(t) \bigr]^2
\nonumber \\
&& 
- {1\over 2} \int_L {d^3\vec k\over (2\pi)^3}\, \phi_{\vec k} \phi_{-\vec k} 
\int_L {d^3\vec k'\over (2\pi)^3}\, F_{k'} \Bigl[ 
\Gamma^{(1)}_4(t;\vec k, -\vec k, \vec k', -\vec k') 
+ \Gamma^{(1)*}_4(t;-\vec k, \vec k, -\vec k', \vec k') 
\Bigr]
\nonumber \\
&& 
+ {2\over 4!} a^3  
\int_L {d^3\vec k_1\over (2\pi)^3} {d^3\vec k_2\over (2\pi)^3}
{d^3\vec k_3\over (2\pi)^3} {d^3\vec k_4\over (2\pi)^3}\, 
(2\pi)^3\, \delta^3(\vec k_1 + \vec k_2 + \vec k_3 + \vec k_4)\, 
\phi_{\vec k_1} \phi_{\vec k_2} \phi_{\vec k_3} \phi_{\vec k_4} 
\nonumber \\
&&\qquad\qquad \times 
\Bigl[ 
F_{k_1} \bigl( \bar\alpha_{k_1} + \bar\alpha_{k_1}^* \bigr)
+ F_{k_2} \bigl( \bar\alpha_{k_2} + \bar\alpha_{k_2}^* \bigr)
+ F_{k_3} \bigl( \bar\alpha_{k_3} + \bar\alpha_{k_3}^* \bigr)
+ F_{k_4} \bigl( \bar\alpha_{k_4} + \bar\alpha_{k_4}^* \bigr)
\Bigr]
\nonumber \\
&&\qquad\qquad \times 
\Bigl[ \Gamma^{(1)}_4(t;\vec k_1, \vec k_2, \vec k_3, \vec k_4) + 
\Gamma^{(1)*}_4(t;-\vec k_1, -\vec k_2, -\vec k_3, -\vec k_4) 
\Bigr]
\nonumber \\
&&\quad
+ {\cal O}(\lambda^2) \biggr\} P_\Omega[\phi_L] . 
\end{eqnarray}
When we choose the coefficient function $F_k$ to reproduce the zeroth order structures in the coarse-grained Liouville equation, and then gather together what remains, we find that
\begin{eqnarray}
i{\partial P_\Omega\over\partial t} 
&=& 
- {i\over 2} \int_{\partial L} {d^3\vec k\over (2\pi)^3}\, 
{1\over a^3}{1\over \alpha_{\Omega,k} + \alpha_{\Omega,k}^*}  
{\delta^2 P_\Omega\over\delta\phi_{\vec k}\delta\phi_{-\vec k}}
+ {i\over 2} \int_L {d^3\vec k\over (2\pi)^3}\, {i\over a^3} 
{\alpha_{\Omega,k} - \alpha_{\Omega,k}^*\over 
\alpha_{\Omega,k} + \alpha_{\Omega,k}^* }
{\delta^2 P_\Omega\over\delta\phi_{\vec k}\delta\phi_{-\vec k}}
\nonumber \\
&&
- {1\over 2} \int_L {d^3\vec k\over (2\pi)^3}\, \phi_{\vec k} \phi_{-\vec k} 
\int_L {d^3\vec k'\over (2\pi)^3}\, 
{1\over\bar\alpha_{k'} + \bar\alpha_{k'}^*}
\Bigl[ \bar\alpha_{k'} \Gamma^{(1)*}_4(-\vec k, \vec k, -\vec k', \vec k') 
\nonumber \\
&&\qquad\qquad\qquad\qquad\qquad\qquad\qquad\qquad\quad
- \bar\alpha_{k'}^* \Gamma^{(1)}_4(\vec k, -\vec k, \vec k', -\vec k') \Bigr]
P_\Omega[\phi_L]
\nonumber \\
&&\quad 
+ {1\over 4!} a^3  
\int_L {d^3\vec k_1\over (2\pi)^3} {d^3\vec k_2\over (2\pi)^3}
{d^3\vec k_3\over (2\pi)^3} {d^3\vec k_4\over (2\pi)^3}\, 
(2\pi)^3\, \delta^3(\vec k_1 + \vec k_2 + \vec k_3 + \vec k_4)\, 
\phi_{\vec k_1} \phi_{\vec k_2} \phi_{\vec k_3} \phi_{\vec k_4} 
\nonumber \\
&&\qquad\qquad \times 
\Bigl[ \bigl[ \bar\alpha_{k_1} + \bar\alpha_{k_2} 
+ \bar\alpha_{k_3} + \bar\alpha_{k_4} \bigr]
\Gamma^{(1)*}_4(-\vec k_1, -\vec k_2, -\vec k_3, -\vec k_4)  
\nonumber \\
&&\qquad\qquad\quad 
- \bigl[ \bar\alpha^*_{k_1} + \bar\alpha^*_{k_2} 
+ \bar\alpha^*_{k_3} + \bar\alpha^*_{k_4} \bigr]
\Gamma^{(1)}_4(\vec k_1, \vec k_2, \vec k_3, \vec k_4) 
\Bigr] P_\Omega[\phi_L]
\nonumber \\
&& 
+ {\cal O}(\lambda^2) . 
\end{eqnarray}

So far we have not used our knowledge of the explicit behaviour of the functions $\alpha_{\Omega,k}$ and $\Gamma_4^{(1)}$.  In this coarsely grained version of the Liouville equation, it is only the long wavelength degrees of freedom that appear --- all of the momenta are in the region well outside the horizon, $k\le \varepsilon aH$.  We are then free to replace $\Gamma_4^{(1)}$ with its asymptotic value, which for a massless field was shown to be
$$
\Gamma_4^{(1)}(\eta;\vec k_1,\vec k_2,\vec k_3,\vec k_4)
= {i\over 3} {\lambda\over H} + {\cal O}(\varepsilon^2) ,
$$
up to corrections suppressed by $k_i^2\eta^2\le \varepsilon^2$, which leaves
\begin{eqnarray}
{\partial P_\Omega\over\partial t} 
&=& 
- {1\over 2} \int_{\partial L} {d^3\vec k\over (2\pi)^3}\, 
{1\over a^3}{1\over \alpha_{\Omega,k} + \alpha_{\Omega,k}^*}  
{\delta^2 P_\Omega\over\delta\phi_{\vec k}\delta\phi_{-\vec k}}
+ {1\over 2} \int_L {d^3\vec k\over (2\pi)^3}\, {i\over a^3} 
{\alpha_{\Omega,k} - \alpha_{\Omega,k}^*\over 
\alpha_{\Omega,k} + \alpha_{\Omega,k}^* }
{\delta^2 P_\Omega\over\delta\phi_{\vec k}\delta\phi_{-\vec k}}
\nonumber \\
&&
+ {1\over 2} \int_L {d^3\vec k\over (2\pi)^3}\, \phi_{\vec k} \phi_{-\vec k} 
\int_L {d^3\vec k'\over (2\pi)^3}\, 
\biggl[ {\lambda\over 3H} \biggr]
P_\Omega[\phi_L]
\nonumber \\
&& 
- {1\over 4!} a^3  
\int_L {d^3\vec k_1\over (2\pi)^3} {d^3\vec k_2\over (2\pi)^3}
{d^3\vec k_3\over (2\pi)^3} {d^3\vec k_4\over (2\pi)^3}\, 
(2\pi)^3\, \delta^3(\vec k_1 + \vec k_2 + \vec k_3 + \vec k_4)\, 
\phi_{\vec k_1} \phi_{\vec k_2} \phi_{\vec k_3} \phi_{\vec k_4} 
\nonumber \\
&&\quad \times 
\Biggl[ 
\bigl( \bar\alpha_{k_1} + \bar\alpha^*_{k_1} \bigr) {\lambda\over 3H}
+ \bigl( \bar\alpha_{k_2} + \bar\alpha^*_{k_2} \bigr) {\lambda\over 3H}
+ \bigl( \bar\alpha_{k_3} + \bar\alpha^*_{k_3} \bigr) {\lambda\over 3H}
+ \bigl( \bar\alpha_{k_4} + \bar\alpha^*_{k_4} \bigr) {\lambda\over 3H}
\Biggr] P_\Omega[\phi_L]
\nonumber \\
&& 
+ {\cal O}(\lambda^2) . 
\end{eqnarray}
The reason that we have written the quartic term in this slightly lengthier form becomes clear when we express it in terms of the coarsely grained potential of the effective theory,
\begin{equation}
{\cal V}_\Omega[\phi_L] = {1\over 4!} \lambda \int_L {d^3\vec k_1\over (2\pi)^3} 
{d^3\vec k_2\over (2\pi)^3} {d^3\vec k_3\over (2\pi)^3} 
{d^3\vec k_4\over (2\pi)^3}\, (2\pi)^3 \delta^3(\vec k_1+\vec k_3+\vec k_3+\vec k_4)
\phi_{\vec k_1}\phi_{\vec k_2}\phi_{\vec k_3}\phi_{\vec k_4} .
\end{equation}
A general Fokker-Planck drift term would have the form
\begin{eqnarray}
&&\hskip-0.375truein
\int_L {d^3\vec k\over (2\pi)^3}\, {\cal D}_k {\delta\over\delta\phi_{\vec k}} \biggl[ {\delta{\cal V}_\Omega\over\delta\phi_{-\vec k}} P_\Omega \biggr] 
\nonumber \\
&=& 
{1\over 2} \lambda \int_L {d^3\vec k\over (2\pi)^3}\, \phi_{\vec k} \phi_{-\vec k} 
\int_L {d^3\vec k'\over (2\pi)^3}\, {\cal D}_{k'} P_\Omega
\nonumber \\
&& 
- {1\over 4!} a^3 \lambda
\int_L {d^3\vec k_1\over (2\pi)^3} {d^3\vec k_2\over (2\pi)^3}
{d^3\vec k_3\over (2\pi)^3} {d^3\vec k_4\over (2\pi)^3}\, 
(2\pi)^3\, \delta^3(\vec k_1 + \vec k_2 + \vec k_3 + \vec k_4)\, 
\phi_{\vec k_1} \phi_{\vec k_2} \phi_{\vec k_3} \phi_{\vec k_4} 
\nonumber \\
&&\qquad\qquad \times 
\Bigl[ (\bar\alpha_{k_1}+\bar\alpha^*_{k_1}) {\cal D}_{k_1}
+ (\bar\alpha_{k_2}+\bar\alpha^*_{k_2}) {\cal D}_{k_2}
+ (\bar\alpha_{k_3}+\bar\alpha^*_{k_3}) {\cal D}_{k_3}
+ (\bar\alpha_{k_4}+\bar\alpha^*_{k_4}) {\cal D}_{k_4}
\Bigr] 
P_\Omega
\nonumber \\
&&
+ \cdots .
\end{eqnarray}
Matching between this general expression and what appears in the Liouville equation of the effective theory, we conclude that ${\cal D}_k=1/3H$ in our theory --- the familiar result.

We have now arrived at the quantum version of the Fokker-Planck equation, evaluated to linear order in the coupling,
\begin{equation}
{\partial P_\Omega\over\partial t} 
= \int_L {d^3\vec k\over (2\pi)^3}\, 
\biggl\{ 
{\cal N}_k {\delta^2P_\Omega\over\delta\phi_{\vec k}\delta\phi_{-\vec k}} 
+ {1\over 3H} {\delta\over\delta\phi_{\vec k}}\biggl[ 
{\delta{\cal V}_\Omega\over\delta\phi_{-\vec k}} P_\Omega \biggr]
\biggr\} + \cdots , 
\end{equation}
where the quantum --- momentum dependent --- noise term is 
\begin{equation}
{\cal N}_k =
- {1\over 2} {1\over a^3}{1\over \alpha_{\Omega,k} + \alpha_{\Omega,k}^*} 
\biggl[ {\partial\over\partial t} \Theta(\varepsilon aH-k) \biggr]
+ {1\over 2} {i\over a^3} 
{\alpha_{\Omega,k} - \alpha_{\Omega,k}^*\over 
\alpha_{\Omega,k} + \alpha_{\Omega,k}^*}\, \Theta(\varepsilon aH-k) 
+ \cdots .
\end{equation}
In the limit where the wavelengths have all been stretched to be much larger than the horizon, it is actually only the first of these terms that determines the leading form of the noise.  Recall that the quantum noise ${\cal N}_k$ and the stochastic noise $N$ are related by 
$$
N = \int_L {d^3\vec k\over (2\pi)^3}\, {\cal N}_k 
$$
Let us evaluate the stochastic noise at leading order in the coupling by replacing $\alpha_{\Omega,k}=\bar\alpha_k+\cdots$ and using the explicit form for $\bar\alpha_k$ in a massless theory,
\begin{equation}
\bar\alpha_k(\eta) = iH{k^2\eta^2\over 1+ik\eta} .
\end{equation}
The stochastic noise coefficient is then found to be 
\begin{eqnarray}
N_0 &=& 
\int {d^3\vec k\over (2\pi)^3}\, \biggl\{
- {1\over 2} {1\over a^3}{1\over \bar\alpha_k + \bar\alpha_k^*} 
\biggl[ {\partial\over\partial t} \Theta(\varepsilon aH-k) \biggr]
+ {1\over 2} {i\over a^3} 
{\bar\alpha_k - \bar\alpha_k^*\over 
\bar\alpha_k + \bar\alpha_k^*}\, \Theta(\varepsilon aH-k) 
\biggr\}
\nonumber \\
&=& 
\varepsilon a {H^4\over 8\pi^2} \int_0^\infty dk\, \biggl\{
{1+k^2\eta^2\over k} \delta(\varepsilon aH-k)
\biggr\}
+ {H^3\eta^2\over 4\pi^2} \int_0^{\varepsilon aH} dk\, k
\nonumber \\
&=& 
{H^3\over 8\pi^2} \bigl( 1+\varepsilon^2 \bigr)
+ {H^3\over 8\pi^2} \varepsilon^2 .  
\end{eqnarray}
In the long wavelength limit, $\varepsilon\ll 1$, we recover the precisely standard noise term for the stochastic Fokker-Planck equation, 
\begin{equation}
{\partial p\over\partial t} 
= {H^3\over 8\pi^2} {\partial^2p\over\partial\varphi^2} 
+ {1\over 3H} {\partial\over\partial\varphi}\biggl( 
{\partial V\over\partial\varphi} p(\varphi) \biggr) ,
\end{equation}
at leading nontrivial order.

\section{Applications and further refinements of the stochastic picture} 

\noindent
We see that the leading form of the quantum version of the Fokker-Planck equation for the effective theory of the long wavelength fluctuations exactly generates the standard Fokker-Planck equation for the stochastic theory.  However, now that we can completely follow the derivation between these two pictures, we can --- as in any effective theory ---refine the basic picture further by evaluating the higher order `corrections' that should appear on the stochastic side by deriving their analogues directly on the quantum side.  For example, we can see that the standard noise and drift,
$$
N(\lambda) = {H^3\over 8\pi^2} + {\cal O}(\lambda) 
\qquad\hbox{and}\qquad
D(\lambda) = {1\over 3H} + {\cal O}(\lambda) ,
$$
are in fact only the first contributions in a perturbative expansion.  What are the forms of the higher order contributions?   Are they also free of late-time divergences?  Do other terms appear in the Fokker-Planck equation?  These last would be the analogues of the higher order operators that appear in the effective Lagrangians in the more familiar applications of effective field theories.

With a means of directly connecting the quantum and stochastic descriptions of the theory, we can --- at least in principle --- explore the behaviour in the late-time limit more fully.  In the static limit of the stochastic theory, the probability function assumes a simple form at leading order in the coupling, e.g.~$p(\varphi)\propto e^{-{\lambda D\over 24N}\varphi^4}$ for the quartic theory.  However, as we mentioned in the introduction, the usual interaction-picture treatments, while consistent with the expectations of the stochastic picture, have late-time divergences that make the approach to this simple, constant, limit difficult to see.  In the Schr\"odinger picture, we have an alternative framework for investigating the behaviour of the quantum theory in this limit.  In particular it would be interesting to learn the explicit time-dependence as the probability function approaches its static limit \cite{progress}.

The technique that we have developed here can also be applied to study the leading behaviour of the stochastic theories associated with other light or massless fields:  multiple interacting scalar fields, gauge fields, or the actual scalar and tensor fluctuations of inflationary theories.  It should be equally instructive to investigate the probability distribution function, $p(t,\varphi(\vec x))$, that is associated with a classical stochastic field.  Such fields are used to describe the long wavelength parts of the $n$-point functions of quantum fields that are evaluated at {\it different\/} spatial positions; these are needed to treat the power spectrum and the non-Gaussianities predicted by inflation.

\acknowledgments

Tereza Vardanyan is grateful for the support of the Department of Energy (DE-FG03-91-ER40682).  We should also like to thank Cliff Burgess for valuable discussions.

\end{document}